\documentclass[a4paper,12pt]{article}

\usepackage[utf8]{inputenc}
\usepackage[dvips]{color,graphicx}
\usepackage{mathrsfs}
\usepackage{amssymb}
\usepackage{amsmath}
\usepackage{slashed}
\usepackage{latexsym}
\usepackage{graphicx}
\usepackage{verbatim}

\newcommand{\be}{\begin{equation}}
\newcommand{\ee}{\end{equation}}
\newcommand{\ba}{\begin{array}}
\newcommand{\ea}{\end{array}}
\newcommand{\bea}{\begin{eqnarray}}
\newcommand{\eea}{\end{eqnarray}}
\newcommand{\balg}{\begin{align}}
\newcommand{\ealg}{\end{align}}
\newcommand{\bit}{\begin{itemize}}
\newcommand{\eit}{\end{itemize}}
\newcommand{\trm}[1]{\textrm{#1}}

\newcommand{\mbb}[1]{\mathbb{#1}}
\newcommand{\msc}[1]{\mathscr{#1}}

\newcommand{\Mpc}{\trm{\Mpc}}
\newcommand{\yr}{\trm{\yr}}
\newcommand{\eV}{\trm{\eV}}

\newcommand{\nn}{\nonumber}

\begin{document}

%%%%%%%%%%%%%%%%%%%%%%%%%%%%%%%%%%%%%%%%%%%%%%%%%%%%%%%%%%%%

\begin{titlepage}
\vspace*{-2cm}
%\flushright{ Here numbers \\  Also here}

\vskip 1.5cm

\begin{center}
{\Large\bf A slant on warped extra dimensions
\vspace{3mm}}
\end{center}
\vskip 0.5cm
\begin{center}
{\large D.~Hernández}$\,^a$~\footnote{dhernand@ictp.it} and 
{\large M.~Sher}$\,^a$~\footnote{mtsher@wm.edu},
\vskip .3cm
%\vskip .cm
$^a\,$ \emph{The Abdus Salam International Centre for Theoretical Physics} \\
\emph{Strada Costiera 11, I-34104 Trieste, Italy}
\\
\vskip .15cm
$^b\,$ \emph{Particle Theory Group, Department of Physics,}\\
\vskip .1cm
\emph{College of William and Mary, Williamsburg, VA 23187, USA}\\
\end{center}
\vskip 0.5cm

\begin{abstract}
We propose an orbifolded, warped, extra dimension scenario in which the visible brane is not parallel to the hidden brane. This leads automatically to Lorentz violation in the visible, four dimensional world.
% that is proportional to the amount of tilting. 
The background solution to the Einstein equations is a function of a parameter that can be identified with the amount of 'tilting' of the brane. The cosmological constant is found to coincide with the classic Randall-Sundrum value to first order in this tilt.  Lorentz violating effects induced in the Standard Model are considered.  We find that the strongest constraint on the tilt comes from determinations of the electron-proton mass ratio in six quasar spectra (four optical and two radio).  %and is given by $5-8 \times 10^{-7}$ per gigaparsec.
Measurements of a third radio source could improve this by an order of magnitude. %  We also consider a more general Lorentz violating background metric.
%new framework of Lorentz violation in scenarios of warped extra dimensions. 
\end{abstract}
\end{titlepage}
\setcounter{footnote}{0}
\vskip2truecm

\newpage

%%%%%%%%%%%%%%%%%%%%%%%%%%%%%%%%%%%%%%%%%%%%%%%%%%%%%%%%%%%%

\section{Introduction}

%%%%%%%%%%%%%%%%%%%%%%%%%%%%%%%%%%%%%%%%%%%%%%%%%%%%%%%%%%%%

Since the natural scale for a fundamental theory is the Planck scale, it has generally been assumed that direct  signals from such a theory would be unobservable, typically being suppressed by powers of $M_W/M_{Pl} \sim 10^{-17}$.    However, if these signals violate Lorentz invariance, then they could be detected in high-precision experiments, some of which\cite{Kostelecky:2008ts}  are even sensitive to the square of $M_W/M_{Pl}$.    As a result, there have been extensive experimental studies of Lorentz violation  (see Ref. \cite{Colladay:1998fq} for a review).    Although theoretical models of Lorentz violation exist, they generally are associated with Planck scale physics \cite{Colladay:1998fq}. Few models involving Lorentz violation in electroweak physics have been presented.     In this paper, we will discuss Lorentz violation in the context of a Randall-Sundrum (RS) class of models which can be tested at the LHC.

The RS scenario is a promising solution to the gauge hierarchy problem \cite{Randall:1999ee, Randall:1999vf}. The model consists of a slice of a five dimensional anti-de Sitter space compactified on an $S_1/Z_2$ orbifold.    At the orbifold fixed points 3D branes are located. The metric is
\be
 ds^2 = e^{-2kr_c|\phi|}\eta_{\mu\nu}dx^\mu dx^\nu + r_c^2 d\phi^2 \label{RS-metric}
\ee
where $x^\mu$ repersent the 4D space-time coordinatesa and $\phi$ is coordinate of the fifth dimension.  The 3D branes consist of a ``hidden" (or UV, or Planck) brane at $\phi=0$ and a ``visible" brane (or IR, or TeV) at $\phi=\pi$.    If one assumes that the Higgs field is confined to the visible brane, then a mass-scale hierarchy of $O(e^{-\pi k r_c})$ is generated, which can give the observed hierarchy for $k r_c \sim 10-11$.

Generalizations of the RS setup have been considered previously. In \cite{Csaki:2000dm}, a warp factor dependent of space and time was considered leading to apparent Lorentz violation and, possibly, violations of causality in the brane. In this paper, we will consider instead the possibility that the radial distance might
% not only vary extremely slowly with time, but also that it might 
have a weak dependence on the spatial coordinates. To linear order, this corresponds to the case of the hidden and visible branes not being exactly parallel. This leads of course to Lorentz violation and in this work, we construct a class of solutions to the 5D Einstein equations that do not posess Lorentz symmetry. 

The deviation from the usual parallel-branes scenario can be parametrized %at linear order
 by a tilt parameter that measures the angle between the branes. Naturally, the solution at this order would predict that the two branes intersect which would be unacceptable. So the tilt parameter must be extremely small, (presumably smaller than an inverse TeV divided by the Hubble Length). But it is nonetheless conceivable that the tilt is nonzero and in this work, we find the current bounds on such a tilt.
%, but the angle is so small that this violation might still be phenomenologically acceptable.  
%Of course, one would not expect the branes to intersect within our Hubble volume, so the tilt angle must be exceptionally small (presumably smaller than an inverse TeV divided by the Hubble length).  But it is conceivable that it is nonzero, and our purpose is to find the current bounds on such a tilt.  The simplest case, which we will consider, is that the distance between the branes varies linearly in space. 

A natural question that emerges is whether the tilted brane is a consistent first order approximation to a full solution of the Einstein equations. We show that indeed this is the case by finding %More generally, we find an extension of the Randall-Sundrum metric
 a generalization of the RS-metric in which the warp factor depends on the spatial coordinates.
%DETAIL THE INTERPRETATION OF THE TILT  (DANI, what do you mean by this?  The above seems clear to me.

There has been much discussion of the mechanism for stabilizing the radial distance in the fifth dimension.   The most studied is the Goldberger-Wise model \cite{Goldberger:1999uk}, in which a radion field is introduced.  In this work, we will not be concerned with the detailed nature of the stabilization mechanism.   We do note that a general study by Steinhardt \cite{Steinhardt:1999eh} suggested a model in which the radial distance can vary extremely slowly with time.

In Section 2, we consider a scenario in which the assumption of 4D Lorentz invariance is dropped, and find the solution to Einstein's equation in the case in which only the leading order term in the tilt parameter is kept.    In Section 3, the phenomenological effects of the Lorentz violation is considered, and the bounds, which arise from studies of quasar spectroscopy, are determined.  In Section 4, we construct a full solution of the Einstein equations that yields a Lorentz violating metric on the brane and relate that result to the phenomenological linear approximation of Section 2. In Section 5, we present our conclusions.

%%%%%%%%%%%%%%%%%%%%%%%%%%%%%%%%%%%%%%%%%%%%%%%%%%%%%%%%%%%%

\section{The tilted metric} \label{sec-lsol}

%%%%%%%%%%%%%%%%%%%%%%%%%%%%%%%%%%%%%%%%%%%%%%%%%%%%%%%%%%%%

%In the original RS scenario the classical action was given as a sum of three terms
%\begin{align}
%S & = S_{grav} + S_{vis} + S_{hid} \label{clas-action}\\
%S_{grav} & = \int d^4x\int_{-\pi}^{\pi}d\phi \sqrt{-G}\{ -\Lambda + 2M^3R\} \\
%S_{vis} & = \int d^4x \sqrt{-g_{vis}}\{ \msc{L}_{vis} - V_{vis} \} \\
%S_{hid} & = \int d^4x \sqrt{-g_{hid}}\{ \msc{L}_{hid} - V_{hid} \} 
%\end{align}
%where
%\be
%g^{vis}_{\mu\nu} = G_{\mu\nu}(x^{\mu},\, \phi=0)\,,\quad g^{hid}_{\mu\nu} = G_{\mu\nu}(x^{\mu},\, \phi=\pi)
%\ee
%{\color{red}(Actually these formulae are very logical and all but they are not what one gets from integrating
%\be
%\int d\phi \sqrt{-G}\delta(\phi)
%\ee
%because the $\mu 5$ components still appear in the determinant $G$. I still don't understand this since Eq.~\eqref{clas-action} is really what one expects if the branes are 4D-Minkowski.)}
The original RS background solution for gravity with full 4D Lorentz invariance is given in Eq.~\eqref{RS-metric}. Key ingredients of the solution were an orbifold structure in the extra dimension and a tuning between the 5D cosmological constant $\Lambda$ and the expectation value of the potential in the branes.

We propose an ansatz for the metric to the Einstein equations of the form
\be
ds^2 = e^{ -2C|\phi| + 2ax|\phi|} \eta_{\mu\nu}dx^\mu dx^\nu +  (R^2 + Dx a)d\phi^2 \,. \label{g-ansatz}
\ee
%where $x_\mu$ refers to the coordinates of the worldly four dimensions while $\phi$ is the additional fifth coordinate. 
The constant $a$ parametrizes the amount of Lorentz violation while $D$ is a constant to be determined. For $\phi =$ const, four-dimensional Lorentz violation is explicit in the $x$-dependence of the diagonal components of the metric. For $a=0$, Eq.~\eqref{g-ansatz}  reduces to the RS solution.
%; that is,
%\be
%f(\phi,x;\, a=0) = e^{-C|\phi|} \,
%\ee
%with $C$ a contant. Hence, we are led to proposing the following ansatz for $f$%:
%\begin{align}
%f & = e^{ -C|\phi| + ax|\phi|} \,.  \label{f-ansatz} % ; \quad C = R\sqrt{ \frac{\Lambda}{24M^3}} 
%f & = e^{ -C(\pi - |t|) + ax|t|} \,; \quad C = R\sqrt{ \frac{\Lambda}{24M^3}}
%\end{align}
%Its physical interpretation is clear. 

The metric in Eq.~\eqref{g-ansatz} is still invariant under the orbifold symmetry $\phi \rightarrow -\phi$. We consider $\phi$ as an angle coordinate with periodic boundary conditions. We take the range of $\phi$ to be $\in [-\pi,\pi]$ with fixed points at 0 and $\pi$ and the space topology is $S^1/\mbb{Z}_2$.  Distances measured along the $\phi$ coordinate for constant $x_\mu$ increase with increasing $Dxa$. This corresponds therefore to a space with two branes at the fixed points % at $\phi=0$ and $\phi=\pi$ 
\emph{tilted} with respect to each other.

The Einstein equations for $\phi$ in the open interval $\phi \in (0,\pi)$ are given by
\be
G_{MN} = -\frac{\Lambda}{4M^3}g_{MN} \label{einstein}
\ee
and can be subsequently solved up to order $a$. To this order the 15 component of the Einstein tensor is different from zero for general $C$ and $D$
\be
G_{15}  = -3\left( 1 +  \frac{CD}{2R^2}\right) \, a \label{15} \,,
\ee
which necessarily yields
\be
2R^2 = -CD \label{constraint} \,.
\ee
%From the 55 component one obtains
%\be
%-\frac{\Lambda}{4M^3} = \frac{6C^2-12Cxa}{R^2 + Dxa} 
%\ee
Applying the constraint in Eq.~\eqref{constraint} to the equations corresponding to the diagonal components in Eq.~\eqref{einstein} we find that the cosmological constant is given by
\be
\Lambda = -\frac{24 C^2M^3}{R^2} \label{cc} \,, 
\ee
exactly as it is found for the case of parallel branes. In other words, there are no contributions to the cosmological constant linear in the tilt. 
%This is non trivial. To order $a$, $\Lambda$ could have had a dependence on the $x$ coordinates. It turns out to be actually \emph{constant} and independent of the tilting of the brane. This is related to another non-trivial fact. That is, 
 Along with Eq.~\eqref{constraint}, Eq.~\eqref{cc} solves the problem completely in terms of the measured $\Lambda$. No adjustable constant remains.
%Finally, it can be checked that the equations 
%\be
%-\frac{\Lambda}{4M^3}g_{\mu\mu} = G_{\mu\mu}
%\ee
%are fulfilled for these values. For completeness and reference I write below the value found for $G_{\mu\nu}$
 %\begin{align}
%G_{\mu\nu} & =  \left[-\frac{6 e^{-2Ct}C^2}{R^2} - \frac{12 e^{-2Ct}C^2 tx}{R^2} \,a \right] \eta_{\mu\nu} \label{munu} 
%\end{align}

We shall see that both features, Eq.~\eqref{constraint} and \eqref{cc}, find a natural explanation when we consider the general solution in Sec.~\ref{sec-gsol}. In particular, for a whole class of $x$-dependent solutions to the Einstein equations, Eq.~\eqref{constraint} is fulfilled automatically and the contributions to the cosmological constant that depend on the coordinates begin at order $a^2$.

There also appears, from the Einstein equations, one for the brane potentials that depends on the second derivatives of the warp factor. We assume for now that it is possible to solve this equation by tuning the brane potentials and we will show that this is indeed the case in Sec.~\ref{sec-gsol}. But first we turn to the phenomenological implications of this scenario.%    Just as in the usual RS scenario, we can introduce some potential in the brane in such a way that it cancels the 5D cosmological constant.

\section{Adding matter}

%%%%%%%%%%%%%%%%%%%%%%%%%%%%%%%%%%%%%%%%%%%%%%%%%%%%%%%%%%%%

There are two versions of the Randall-Sundrum model:  either  the Standard Model fields are on the brane or in the bulk.   We will take the Higgs field to be confined to the visible brane at $\phi=\pi$.     First, we will consider gauge bosons and fermions to be on the brane, and then will consider them to be in the bulk.  In the latter case, the fermion mass hierarchy problem can be solved geographically, by localizing the fermions at different locations in the bulk \cite{Gherghetta:2000qt, Grossman:1999ra, Huber:2000ie, Huber:2003tu}.

The action for the Higgs field is the action reduced to that brane and takes the form
\begin{align}
S_H & = \int d^4x \sqrt{-g_{vis}}\big[ g^{\mu\nu}_{vis} D_\mu H^\dagger D_\nu H - \lambda(H^\dagger H - v_0^2)^2 \big] \\
& = \int d^4x \big[ e^{-2\pi(C-ax)}\eta^{\mu\nu}_{vis} D_\mu H^\dagger D_\nu H - \lambda e^{-4\pi(C-ax)}(H^\dagger H - v_0^2)^2 \big]
\end{align}
Redefine
\be
H \rightarrow e^{\pi(C-ax)}H'
\ee
Then,
\be
D_\mu H = e^{\pi(C-ax)} \big( D_0 H'\,, \;\; -\pi aH' + D_1 H'\,, \;\; D_2 H' \,,\;\; D_3 H' \big)
\ee
and hence
\begin{align}
S_{H'} & = \int d^4x \big[ \eta^{\mu\nu} D_\mu H'^\dagger D_\nu H' - \eta^{\mu\nu}( a_\mu H'^\dagger D_\nu H' + \trm{h.c.} ) - \nn \\
& \quad\quad\quad - \lambda(H'^\dagger H' - e^{-2\pi(C-ax)}v_0^2 )^2 \big]
\end{align}
with $a_\mu = (0,\pi a,0,0)$. Therefore, in the pure Higgs sector, Lorentz violation appears in two ways, one as a $(k_\phi)^\mu H'^\dagger D_\mu H'$ term \cite{Colladay:1998fq} and two, as a space varying Higgs vev.

The  $(k_\phi)^\mu H'^\dagger D_\mu H'$ term can easily be seen to have a negligible effect.   The typical value of $a_\mu$ cannot be bigger than an inverse gigaparsec, or else the branes will intersect within our horizon.   But a gigaparsec is approximately $10^{40}$ inverse GeV, and  the experimental upper bound on the $(k_\phi)^\mu$ term is no greater than $10^{-27}$ GeV.  Thus experiments are  many orders of magnitude shy of being sensitive to that term.   Instead, the primary effect will come from the space varying vev.   We first consider a spatially varying $a_\mu$, and then a time varying $a_\mu$.

%%%%%%%%%%%%%%%%%%%%%%%%%%%%%%%%%%%%%%%%

\subsection{Fermion and gauge fields on the brane}

%%%%%%%%%%%%%%%%%%%%%%%%%%%%%%%%%%%%%%%%

If the fermions and gauge fields are on the brane, the Yukawa couplings are not affect by the tilt and thus the only effect is due to the variation of the Higgs vev.
To leading order in $a$, the vev of the Higgs field varies as $v_0 (1 + \pi a x)$.    This will mean that the quark and lepton masses scale as $m_0 (1 + \pi a x)$.      In Ref. \cite{Agrawal:1997gf, Agrawal:1998xa}, anthropic constraints on the allowed variation of $v$ were considered.   Later, in Ref. \cite{Jeltema:1999na}, it was noted that the strongest bound comes from considering the triple-alpha process in stars---if $v$ is less than $90\%$ of Standard Model value, then the carbon resonance in the triple alpha process is shifted and carbon production does not occur.   Thus, the existence of supernovae at large distances (and in all directions) would immediately imply a bound of roughly $ \pi a  <  0.1/(3 {\rm Mpc})$.      It should be noted that although the quark masses scale linearly as the VEV of the Higgs field, the nucleon mass does not, since the mass is set primarily by the QCD scale.  In Ref. \cite{Agrawal:1997gf, Agrawal:1998xa}, it was noted that the QCD scale is also sensitive to the VEV of the Higgs field through quark threshold effects, and they estimated that the scale varies as $(v/v_0)^\zeta$, where $\zeta$ varies between $0.25$ and $0.3$--we will take it be $0.25$ here.  Thus, the proton mass scales as $(1 + 0.25\pi a x)$.

Stronger bounds can be obtained from the cosmic microwave background radiation (CMBR).     The temperature at the surface of last scattering depends on the binding energy of hydrogen, which varies linearly with the electron mass.   Thus one would expect the temperature to vary as a dipole distribution.   Unfortunately, there already is a dipole distribution in the CMBR due to the Earth's motion, and there does not appear to be a simple way to separate these effects.  Nonetheless, unless the Earth's motion happens to line up in opposite direction as the tilt, a bound can be obtained.   Thus, of the three spatial components of $a$, two linear combinations can be bounded (although we don't know what these combinations are since we don't know the direction of Earth's motion).  The dipole is $\Delta T/T = 1.2 \times 10^{-3}$ and thus one would obtain a bound of $\pi a < 0.0012/R$, where $R$ is the Hubble radius.   With fine-tuning, of course, this bound could be completely evaded.

The strongest bounds, which cannot be eliminated by fine-tuning, come from precise spectral lines of distant objects.     Again, a shift in the electron mass can't be distinguished from a red-shift or blue-shift, however observations of numerous spectral lines will be sensitive to the electron-proton mass ratio, $\delta \mu/\mu$ independently of the overall shift.    This will then scale as $-.075\pi a x$.  We will choose our x-axis to maximize the upper bound on $a$.

There have been numerous studies of the electron-proton mass ratio for distant objects.   A comprehensive analysis can be found in the recent work of Malec et al.\cite{Malec:2010xv}.    They used the Keck telescope to study molecular transitions at redshifts of $z > 2$ in the direction of bright background quasars.   At those redshifts, the Lyman lines of hydrogen move into the region where they penetrate the atmosphere and can be detected on the ground.    In one particular direction, they find $\delta\mu/\mu = (5.6 \pm 5.5_{stat} \pm 2.9_{sys})\times 10^-6$ for a redshift of $z=2.059$.   Of course, the tilt direction could be perpendicular to that, and thus several measurements in different directions are needed, and they summarize the other extra-galactic measurements of the ratio.   In particular, they mention two limits from radio studies of $NH_3$ which give bounds roughly an order of magnitude stronger, but at redshifts somewhat less than $1.0$.

The results are listed in Table I.   We have listed the measurement (with errors added in quadrature), the redshift and the distance in comoving coordinates.   For each object, the first four numbers of the designation give the right ascension in hours and minutes, and the sign plus the remaining numbers give the declination in degrees (two digits plus a decimal point).

\begin{table}[ht]
\centering
\begin{tabular}{lclclclclcl}
\hline
 Expt. & authors & Object  & Redshift & Dist.  & bound ($\times 10^{-6}$) \\\hline\hline
 $H_2/HD {\rm Keck}$ & Malec\cite{Malec:2010xv} &  J2123-0050  & 2.059 & 5.26& $5.6 \pm 6.2 $\\\hline
$H_2/VLT$ & King \cite{king} & Q0405-443 &  2.595 &5.94&  $10.1\pm 6.2$\\\hline
 & & Q-0347-383  & 3.025 & 6.38 & $8.2 \pm 7.4$\\\hline
 & & Q-0528-250& 2.811 &  6.17& $-1.4 \pm 3.9$\\\hline
$NH_3$ &Murphy \cite{murphy}& B0218+357 &  0.68 & 2.45& $0.74 \pm 0.89 $\\\hline
$NH_3$& Henkel \cite{henkel}& PKS1830-211 & 0.89 & 3.03& $0.08 \pm 0.47$
\\\hline
\end{tabular}
\caption{Bounds on the electron-proton mass difference from six observations of distant spectra.  The experiments are given, along with the object, the redshift, the distance away in comoving coordinates in units of gigaparsecs, and the bound.}
\end{table}

Note that if there were only two measurements, then there would be no bound, since one could choose the x-axis to lie on the perpendicular to the plane consisting of the earth plus these two objects, and thus $x$ would vanish.  But with six objects, one can get a bound.  Thus, even though two of the limits are much tighter, they will not control the bound--a third is needed.    For a given value of $a,\theta,\phi$, we find the values in which the $\chi$-square is minimized.   One can easily see from the table that the $\chi$-square if $a=0$ is $5.5$.   As $a$ increases, the $\chi$-square drops, passing a value of $4.66$ at $0.75\pi a = 0.15\times 10^{-6} {\rm Gpc}^{-1}$, reaching a minimum value of $2.66$ at $0.75\pi a = 0.94\times 10^{-6}{\rm Gpc}^{-1}$, and then increases, reaching $4.66$ at $0.75\pi a = 1.76  \times 10^{-6}{\rm Gpc}^{-1}$.   Therefore, we conclude that the value of $0.75\pi a$ that one obtains from the data, to two standard deviations, is $(0.94 \pm 0.8)  \times 10^{-6}{\rm Gpc}^{-1}$, which leads to a $2\sigma$ upper bound on $a$ of $7.5\times 10^{-7}{\rm Gpc}^{-1}$.

Note that there are two very precise measurements, and yet a third is needed to control the bound.  Thus, a third radio limit from $NH_3$ would improve the bound by close to an order of magnitude.

In the above, we have considered the case in which $a$ is spacelike.  If it is timelike, then the orientation is irrelevant.  This precise issue was considered by Henkel et al.\cite{henkel}, who found that over the past 7.0 Gyr, the $3\sigma$ upper limit on the  time variation of $\mu$ is $2.0\times 10^{-16}\ {\rm yr.}^{-1}$.   Multiplying by the speed of light,  these results can be easily translated into our parameters, and gives a $3\sigma$ upper bound on $a$ of $1.4\times 10^{-7}{\rm Gpc}^{-1}$.

%%%%%%%%%%%%%%%%%%%%%%%%%%%%%%%%%%%%%%%%

\subsection{Fermion fields in the bulk}

%%%%%%%%%%%%%%%%%%%%%%%%%%%%%%%%%%%%%%%%

If the fermions are in the bulk, the electron-proton mass ratio is not just given by the changing VEV, but also by the Yukawa couplings.  This affects not only the electron mass, but also the proton mass through the change in $\Lambda_{QCD}$ through threshold effects, as noted by Agrawal et al.\cite{Agrawal:1997gf,Agrawal:1998xa}.    The effect is to change the $0.75$ in the above analysis.

Recall the conventional treatment of fermions in the bulk.    As shown in Ref. \cite{Gherghetta:2000qt}, the fermion wave equation is given by
\be
[e^{2\sigma}\eta^{\mu\nu}\partial_\mu\partial_\nu + e^\sigma\partial_5(e^{-\sigma}\partial_5)-M^2]e^{-2\sigma}\Psi_{L,R} = 0 \label{f-wave}
\ee
where  $\sigma=k|r|$ and 
\be
M^2 = c(c\pm 1)k^2 \mp c\sigma^{\prime\prime}
\ee
with the $\pm$ refering to the left and right components. $c$ is defined as the proportionality constant in $m_\Psi = c\sigma'$ where $m_\Psi$ is the 5D fermion mass.   The solution of Eq.~\eqref{f-wave} can be written in terms of Bessel functions.  As shown in Ref. \cite{Gherghetta:2000qt}, the Yukawa coupling of the Standard Model fermions to the Higgs is given by the overlap of the zero mode fermions with the Higgs field on the visible brane, and is given by
(suppressing flavor indices)
\be
g_Y = {g_Y^5 k\over N_LN_R}e^{(1-c_L-c_R)\pi k R}
\ee
where
\be
{1\over N_L^2} \equiv {1/2 - c_L \over e^{(1-2c_L)\pi kR}-1}
\ee
and similarly for $N_R$.    Here, $g_Y^5$ is the 5D Yukawa coupling, and $c_L,c_R$ are the left and right handed 5D mass terms.  Note that this leads to exponentially suppressed Yukawa couplings for $c_L$ and $c_R$ somewhat larger than $1/2$.    Note that each fermion mass (assuming $O(1)$ 5D Yukawa couplings) can be given in terms of $c_L$ and $c_R$, but the individual values of these mass parameters are not determined.  Gherghetta and Pomarol \cite{Gherghetta:2000qt} consider two limits:  $c_L = c_R$ and $c_R = 1/2$, which simplifies the expression for the Yukawa couplings.  

In the tilted brane scenario, one should write the wave equation in the new metric, find the zero modes and calculate the overlap on the visible brain.   Unfortunately, with an off-diagonal metric, the wave equation is non-separable, making an analytic solution impossible.   Since the ratio of $c_L$ to $c_R$ is arbitrary, which will lead to substantial uncertainty in our results, we can make an approximation.   At a position $x$, the value of $R$ is changed by $kR\rightarrow kR + ax$.  We will thus replace $kR$ in the above by $kR+ ax$, and obtain the Yukawa coupling as a function of $x$.  Comparing to the Yukawa coupling at $x=0$, the unknown $g_Y^5$ will cancel out.   In effect, this approximation is replacing the tilted brane at position $x$ with a parallel brane at the same value of $R$.  Since $a$ is very small, this approximation seems reasonable.

For the $c_L = c_R$ case,  one finds that $\Delta g_Y/g_Y = 2 (c_L - 1/2) \pi a x$.   To give the correct electron mass, $c_L = 0.63$, so this gives $0.26 \pi a x$ (note that the effects of the other fermions are much smaller).    This has the effect of changing the $0.75$ in the previous section to $1.01$.  For the $c_R = 1/2$ case, it gives $0.13 \pi a x$, changing the $0.75$ to $0.88$.     Since we don't know the ratio, all we can say is that the bounds in the above section are tightened by a factor ranging from $15 - 30$ percent.

%%%%%%%%%%%%%%%%%%%%%%%%%%%%%%%%%%%%%%%%%%%%%%%%%%%%%%%%%%%%

\section{General Lorentz violating background metric} \label{sec-gsol}

%%%%%%%%%%%%%%%%%%%%%%%%%%%%%%%%%%%%%%%%%%%%%%%%%%%%%%%%%%%%

In this section we will construct a more general Lorentz-violating, Randall-Sundrum-like background metric with an $x$ dependence in the warp factor.  Consider then the Einstein equation
%In order to do that we begin by constructing a 
%We construct in what follows a general Lorentz violating solution to the Einstein equation
\be
G_{MN} = -\frac{\Lambda(x)}{4M^3}g_{MN} \label{eeq}
\ee
where we have allowed for the possibility of an $x$-coordinate dependence on the cosmological ``constant''. %For the case of constant $\Lambda$ we will find 
%A general Lorentz violating solution to the Einstein equations that includes the metric analyzed in the text as a particular case can be constructed as follows. 
We begin by noticing that the ansatz
\be
ds^2 = e^{2k(x,|\phi|)}\eta_{\mu\nu}dx^\mu dx^\nu + R^2_c\dot{k}^2(x,|\phi|)d\phi^2 \label{metric-ans}
\ee
yields a diagonal Einstein tensor $G_{MN}$.  Here
\be
\dot{k} = \frac{\partial k}{\partial \phi} \;,\quad k' = \frac{\partial k}{\partial x} \, \,.
\ee
From now on, whenever $k$ or its derivatives appear they are understood to represent the functions with the second argument evaluated in $|\phi|$ unless stated otherwise.
%\tcr{and we assume for the time being that $k(x,\phi)$ is $\mcl{C}^\infty$. As a matter of fact, strictly speaking this ansatz does not constitute a solution of the Einstein equations for the interval $\phi \in [-\pi,\,\pi$]. In particular the orbifold boundary conditions imply that $k(x,\phi)$ must be an even function of $\phi$ will yields $\dot{k}(x,\phi=0) = 0$. Thus, the metric in Eq.~\eqref{metric-ans} is not invertible at $\phi=0$. We will see that this subtlety will be fixed when we impose non-differentiability at the brane. In the other hand, if we don't impose the orbifold boundary conditions, the ansatz is perfectly defined and coincides with the linear order solution found in Sec.~\ref{sec-lsol} for $\phi \neq 0$ and
%\be
%k(x,\phi) = -C\phi\left( 1 - \frac{ax}{C} \right) \,,\quad R_c = \frac{R}{C} \,.
%\ee
%}
%and $F(\phi)$ is an arbitrary function of $\phi$. 
%Notice however that Eq.~\eqref{metric-ans} as it is  does not reduce to the RS ansatz. In particular, if $k(x,\phi)$ were proportional to $|\phi|$, the derivative $\dot{k}$ would be undefined at $\phi=0$. We thus impose that $k(x,\phi)$ hshould be  

The expression for $\Lambda(x)$ can be found from the 55 component of Eq.~\eqref{eeq} % we obtain the expression for $\Lambda$
\be
\Lambda(x) = -\frac{24M^3}{R^2_c} - 12M^3e^{-2k}(k'^2 + k'') \,. \label{Lx}
\ee
$k(x,|\phi|)$ can be expanded as a power series in a small tilt $ax$
\be
k(x,\phi) = k_0(|\phi|) + axk_1(|\phi|) + (ax)^2k_2(|\phi|) + \dots
\ee
and from Eq.~\eqref{Lx} it is clear that the cosmological constant is only corrected at second order in the tilt $a$. This explains why the linear approximation gives a cosmological constant which is $x$-independent.
%Notice that the first order solution in the text corresponds to the particular case
%\be
%k = -C|\phi|\left( 1 - \frac{ax}{C}\right) \; ,\quad F = \frac{1}{C} \,.
%\ee
%As it was remarked in the text, this solution is necessarily approximate since $\dot{k}$ is discontinuous. The second term in Eq.~\eqref{Lx} yields the $x$-dependence of the cosmological constant. As it is expected for this solution, it appears only at second order in $a$.

\vspace{.3cm}
There are two more independent equations appearing from Eq.~\eqref{eeq}. Plugging in  Eq.~\eqref{Lx} we find from the 11 component that it must be
\be
k' \dot{k'} = k'' \dot{k}  \label{k-r}\,,
\ee
%, we find  (plugging in) 
with the trivial solution $k'=0$ which corresponds to Randall-Sundrum. However, for $k' \neq 0$ another solution appears:
\be
k' = \dot{k}F(|\phi|) \label{k-rel} \,.
\ee
%Here $F(\phi)$ is in principle an arbitrary function of $\phi$.
%We will focus on the latter from now on. 
%From the orbifold conditions at $\phi=0$ it is clear that % $k'$ and $\dot{k}$ have opposite orbifold parities and 
%$F$ must also be a function of $|\phi|$.}  
Notice that any function of the form $k(x,\phi) = h(e^{\alpha x}\kappa(\phi))$
with $\alpha$ a constant, is a solution to Eq.~\eqref{k-rel} with $F(\phi) = \alpha\kappa(\phi)/\dot{\kappa}(\phi)$ \,.

The 00 component of the metric provides the last equation:
\be
%\frac{6e^k(e^k-1)}{R^2F^2} + 
2k'^2+k'' - \frac{k'\dot{k'}+\dot{k}''}{\dot{k}} = 0
\ee
or, using Eq.~\eqref{k-rel},
\be
%\frac{6e^k(e^k-1)}{R^2F^2} + k'^2 + 2k'' - 3e^{-k}(k'^2+k'') + \frac{k'k''+ k'''}{k'} = 0 \,.
2k'^3 - k''' = 0 \,. \label{final-eq}
\ee
%Changing variables to $g$ we finally find the simple form
%\be
%\frac{6g(g-1)}{R^2F^2} + \frac{g'''}{g'} - \frac{3g''}{g^2} = 0 \,. \label{g-eq}
%\ee
Consider now this as an ordinary differential equation. With proper boundary conditions, there is a unique solution to it, $k_0(x)$. Dropping one of the boundary conditions we can find a full orbit of solutions depending on one constant $k_0^C = k_0(x+C)$ since $x$ doesn't appear explicitly in Eq.~\eqref{final-eq}. Now, it is always possible to define a function $k'$ by the relation $k'_0(e^x) = k_0(x)$. Hence $k_0'^C = k_0'(e^{x+C})=k_0'(e^x C')$ is a solution of  Eq.~\eqref{final-eq}. But, we can substitute $C'$ by an arbitrary function of $\phi$ and it will still solve  Eq.~\eqref{final-eq}. Hence we conclude that there is a solution of  Eq.~\eqref{final-eq} that fulfills the condition that it is a function of $e^{\alpha x}\kappa(\phi)$ and hence is also a solution to Eq.~\eqref{k-rel}.% Finally, in order to fulfill the orbifold condition at the fixed point $\phi=0$ we require that $k_0'$ should be an even function of $\phi$. 
With this, the proof that a Lorentz violating background metric exists to all orders is complete and it is also clear its relation to the solution discussed in the text.   

Eq.~\eqref{final-eq} can be solved analytically. We obtain
%\be
%k' = (-c_1)^{1/4} \trm{Sn}_{-1}\left[ (-c_1)^{1/4}(x+c_2)\right] \label{k'}
%\ee
\be
k = \trm{Log}\left\{ e^{i\pi/4}\left[-i\trm{Cn}\Big(e^{3i\pi/4}c_1(x+c_2)\Big) + \trm{Dn}\Big(e^{3i\pi/4}c_1(x+c_2)\Big) \right] \right\} + c_3 \label{k}
\ee
where $\trm{Cn}(x)$ and $\trm{Dn}(x)$ are the Jacobi elliptic functions defined through the elliptic integral
\be
u(\eta) = \int_0^\eta \frac{d\theta}{\sqrt{1+\sin^2\theta}}
\ee
as $\trm{Cn}(u(\eta)) = \cos \eta$ and $\trm{Dn}(u(\eta)) = \sqrt{1+\sin^2\eta}$ (we will also need in a moment $\trm{Sn} (u(\eta)) = \sin \eta$). Substituting the constants $c_i$ for functions of $\phi$ in Eq.~\eqref{k} one has the general solution of Eq.~\eqref{final-eq}. However, in order to make it consistent with Eq.~\eqref{k-r} only $c_2$ \emph{or} $c_3$ can be made nonconstant functions of $|\phi|$ as explained above (not both at the same time!). %(DANI--I don't understand these last two sentences--what does "promoted" mean).

Randall-Sundrum corresponds to the case in which $c_1 = c_2 = 0$, $c_3 = -C|\phi|$ and $R_c = R/C$. On the other hand, for $c_3=0$, $c_2(\phi) \neq 0$ one obtains the Lorentz-violating solution to the Einstein equations.

It is possible to expand Eq.~\eqref{k} both in $x$ and in $\phi$ to find the linear approximation discussed in Sec.~\ref{sec-lsol}. As an example, the constant $C$ from Eq.~\eqref{g-ansatz} is expressed in terms of the function $c_2(\phi)\simeq \alpha + \beta \phi + o(\phi^2)$ and the constant $c_1$ by
\be
C = - e^{3\pi i/4} \beta c_1\trm{Sn}(e^{3\pi i/4}\alpha c_1) \,.
%[\trm{Cn}_{-1}(e^{3\pi i/4}\alpha c_1) + i\,\trm{Dn}_{-1}(e^{3\pi i/4}\alpha c_1) ]}{}
\ee
%where $c_2(\phi) $. 
Finally, it can be checked that the equations on the branes turn out to be, for this more general background, exactly the same as in the RS case. For the brane action
\be
S_{\phi=0(\pi)} = \int d^4x \sqrt{g_{\phi=0(\pi)}} (\msc{L}_{\phi=0(\pi)} -V_{\phi=0(\pi)})
\ee
One finds
\be
%\frac{3 \dot{k}(x,|\phi|)}{R\dot{k}(x,\phi)} = 
\frac{3}{R_c}\delta(\phi) = \frac{V_{\phi=0}}{4M^3}\delta(\phi) \,.
\ee
and a similar one for the brane at $\phi=\pi$. The potential on the brane does not depend on coordinate $x$.

\section{Conclusions}

Since experiments involving Lorentz violation can be extremely precise, it has been the subject of numerous phenomenological studies over the past decades.   Yet most such studies are parametric, in which specific Lorentz violating terms which could exist in a Lagrangian are analyzed and discussed.   There are relatively few plausible models in which Lorentz violation emerges.

In this paper, we have considered a model in which the two branes of the Randall-Sundrum model are not exactly parallel.    Although we do not have a specific motivation for this model, the possibility that the branes could intersect well outside of our Hubble radius could have fascinating cosmological consequences.  Our aim in this paper is to study the phenomenological implications of a small angle between the branes.

In Section 2, we present a simple linear ansatz for the metric, in which the distance between the branes varies linearly with position (along some spacelike direction).    It is shown that the ansatz naturally leads to a bulk cosmological constant which is independent of position.    In Section 3, the phenomenological implications are studied.   The strongest bounds come from studies of the electron-proton mass ratio in distant astrophysical objects.    We find that the distance between the branes cannot vary by more than 
a factor of $7.5\times 10^{-7} {\rm Gpc}^{-1} =  5\times 10^{-51} {\rm TeV}$.      Since the most precise measurements of the electron-proton mass ratio come from two observations (and the tilt direction could, in principle, be perpendicular to the plane of the Earth and those objects), a third observation would improve the bound by up to an order of magnitude.  If one believes that the factor should be a power of the weak scale to the Planck scale, or $10^{-17n}$ where $n$ is an integer, then the bound is very close to $n=3$.      If the Standard Model fields are in the bulk, the bounds are strengthened by between 15 and 30 percent.

Finally, we obtained a full solution to the Einstein equations that generalizes RS and that, in the general case, yields Lorentz violation in the branes. The linear expansion of this solution corresponds to the phenomenological linear tilt analized before. This puts the approximation on firm ground. Inside the Hubble radius, the RS solution to the hierarchy problem works as before. %more general solution, without making the linear approximation.  
 %A model in which the metric is infinitely differentiable is studied, and a general solution is found.  
%The conditions for this to lead to a constant bulk cosmological constant are given.  
%We then use this solution to show that a solution can exist for Randall-Sundrum type models which are non-differentiable at the origin and require brane potentials.
\vskip 1cm
{\noindent{\bf Acknowledgments}}

We are thankful to Josh Erlich for many useful discussions and suggestions, and to Amy Barger, David Sher and José Roberto Vidal for useful conversations. MS is very grateful to the Universidad Autonoma de Madrid for its hospitality where this work was begun, and acknowledges support from the National Science Foundation grant NSF-PHY-0757481. DH acknowledges partial financial support from the Spanish Ministry of Education and Science (MEC) through FPU grant AP20053603. 

%%%%%%%%%%%%%%%%%%%%%%%%%%%%%%%%%%%%%%%%%%%%%%%%%%%%%%%%%%%%

\end{document}